\documentclass[aps,prc,twocolumn,showpacs,superscriptaddress,floatfix,nofootinbib]{revtex4-1}
\usepackage{amssymb}
\usepackage{amsmath}
\usepackage{graphicx}

\newcommand*{\scale}
{{\text{sc}}}
\newcommand*{\threshold}
{{\text{th}}}
\newcommand*{\ystrut}
[1]{\raisebox{#1}{\mbox{}}}
\newcommand*{\mF}
{\mathcal{F}}
\allowdisplaybreaks

\begin{document}

\title{\boldmath Analysis of recent CLAS data on $\Sigma ^{\ast }(1385)$
photoproduction off a neutron target}
\author{Xiao-Yun Wang}
\thanks{xywang@impcas.ac.cn}
\affiliation{Institute of Modern Physics, Chinese Academy of Sciences, Lanzhou
730000, China} \affiliation{University of Chinese Academy of Sciences,
Beijing 100049, China}
\affiliation{Research Center for Hadron and CSR Physics, Institute of Modern Physics
of CAS and Lanzhou University, Lanzhou 730000, China}
\author{Jun He}
\thanks{Corresponding author: junhe@impcas.ac.cn}
\affiliation{Institute of Modern Physics, Chinese Academy of Sciences, Lanzhou
730000, China}
\affiliation{Research Center for Hadron and CSR Physics, Institute of Modern Physics
of CAS and Lanzhou University, Lanzhou 730000, China}
\affiliation{State Key Laboratory of Theoretical Physics, Institute of Theoretical
Physics, Chinese Academy of Sciences, Beijing 100190, China}
\author{Helmut Haberzettl}
\thanks{helmut.haberzettl@gwu.edu}
\affiliation{Institute for Nuclear Studies and Department of Physics, The
George Washington University, Washington, DC 20052, USA}
\begin{abstract}
Based on recent experimental data obtained by the CLAS Collaboration, the $%
\Sigma(1385)$ photoproduction off a neutron target at laboratory photon
energies $E_\gamma$ up to 2.5\,GeV is investigated in an effective
Lagrangian approach including $s$-, $u$-, and $t$-channel Born-term
contributions. The present calculation does not take into account any
explicit $s$-channel baryon-resonance contributions, however, in the spirit
of duality, we include $t$-channel exchanges of mesonic Regge trajectories.
The onset of the Regge regime is controlled by smoothly interpolating
between Feynman-type single-meson exchanges and full-fledged
Regge-trajectory exchanges. Gauge invariance broken by the Regge treatment
is fully restored by introducing contact-type interaction currents that
result from the implementation of \textit{local} gauge invariance in terms
of generalized Ward-Takahashi identities. The cross sections for the $\gamma
n\to K^+\Sigma^*(1385)^-$ reaction are calculated and compared with
experimental results from the CLAS and LEPS collaborations. Despite its
simplicity, the present theoretical approach provides a good description of
the main features of the data. However, the parameters fitted to the data
show that the gauge-invariance-restoring contact term plays a large role
which may point to large contributions from final-state interactions.
\end{abstract}

\maketitle

\section{Introduction}

Over the last few decades, great strides have been made in baryon spectroscopy,
in large parts thanks to the high-quality photoproduction data obtained at
electromagnetic facilities such as JLab, MAMI, ELSA, SPring-8, BEPC, and
others.  Kaon photoproduction, in particular, with a strange (ground-state) $\Sigma$
baryon, has been extensively studied. However, there exist only a limited
number of studies of kaon photoproduction with a strange $\Sigma^{\ast}(1385)
$ ($\equiv \Sigma^{\ast}$) baryon resonance~\cite%
{ca67,re69,mn08,CLAS13,sc08,yo08,gao10,he14}. In this respect, studies of
the photoproduction of the $\Sigma^{\ast}(1385)$ off a neutron are
particularly scarce both experimentally and theoretically. Recently, the
JLab CLAS Collaboration released preliminary experimental data for the $%
\gamma n\rightarrow K^{+}\Sigma^{\ast}(1385)^{-}$ process~\cite{clas14},
where it was found that the differential cross sections of the CLAS
experiment are in agreement with the published LEPS Collaboration results \cite{leps09}.
The present work provides an exploratory study of the dominant mechanisms
that provide an understanding of the $\gamma n\rightarrow K^{+}\Sigma ^{\ast
}(1385)^{-}$ reaction, based on these two data sets.

To this end we adopt here an effective Lagrangian approach in terms of
standard $s$-, $u$-, and $t$-channel exchanges, similar to the studies of $%
\Lambda(1520)$ and $\Sigma(1385)$ photoproductions off the proton~\cite%
{yo08,gao10,he14,He:2012ud,Wang:2014jxb,He:2014gga}. At high energies,
however, a more economical approach may be furnished by a phenomenological
Regge treatment~\cite{Collins,DDLN,Gribov}. Hence, to be able to adapt the
standard effective-Lagrangian description and provide a transition into the
Regge regime at higher energies, we adopt here a method that smoothly
interpolates between Feynman-type low-energy single-meson $t$-channel
exchanges and a full-fledged high-energy Regge treatment. Such a hybrid
approach was seen to be quite successful in reproducing the experimental
data in Refs.~\cite{he14,He:2012ud,Wang:2014jxb,He:2014gga}. The present
treatment, however, is different from previous approaches in two important
aspects.

First, in the present work, we will
\textit{not} include any baryon resonances in the $s$ channel. We do not do
so because the few CLAS and LEPS data points available do not exhibit any
rapid variation with energy and angle~\cite{clas14,leps09}, which suggests
that a calculation that concentrates on the major background mechanisms
should be capable of capturing the main features of the data. Moreover,
duality suggests that a full set of $t$-channel exchanges is equivalent to a
full set of $s$-channel resonances~\cite{Collins,DDLN,Gribov}. Taking into
account both, therefore, would correspond to double counting. While we do
not suggest that the somewhat simplified Reggeized $t$-channel treatment
described below corresponds to a true \textit{full} set of $t$-channel
exchanges in the sense of duality, we want to explore this avenue here to
see whether one can describe the dominant features of the data without
explicit $s$-channel resonances. Such an exploratory investigation may be
thought of as a sort of ``poor-man's duality" treatment.

Second, to repair gauge invariance broken by the implementation of $t$%
-channel Regge exchanges, we will employ here the method we recently proposed
~\cite{Haberzettl:2015exa} which is based on requiring the
\textit{off-shell} photoproduction current $\mathcal{M}^\mu$ to satisfy the
generalized Ward-Takahashi identities that follow from consistently imposing
\textit{local} gauge invariance at the microscopic level~\cite%
{Kazes1959,Haberzettl:1997jg}. The procedure involves constructing a minimal
\textit{contact}-type interaction current utilizing Regge-trajectory
exchanges, similar to what is proposed in Ref.~\cite{hh06} for ordinary
Feynman-type single-hadron exchanges. The complete on-shell production
current thus obtained satisfies the necessary (global) gauge-invariance
condition $k_\mu\mathcal{M}^\mu=0$ as a matter of course (with $k$ being the
photon four-momentum). As far as local gauge invariance is concerned, the
procedure of Ref.~\cite{Haberzettl:2015exa} utilized here is dynamically
complete. It is markedly different from the often-used prescription proposed
in Ref.~\cite{GLV97}; while the corresponding \textit{ad hoc} recipe does
indeed produce a globally gauge-invariant production current, it is without
dynamical foundation, however.

The paper is organized as follows. In the subsequent Sec.~\ref{sec:formalism}%
, we present the formalism and main ingredients used for describing the reaction
$\gamma n\rightarrow K^{+}\Sigma^{\ast}(1385)^{-}$. The details of the
interpolating Regge treatment and restoration of the local gauge invariance
are also presented there. Numerical results are discussed in Sec.~\ref%
{sec:results}, followed by a brief summary in Sec.~\ref{sec:summary}.

\section{Formalism}

\label{sec:formalism}

\begin{figure}[tbph]
\begin{center}
\includegraphics[scale=0.48]{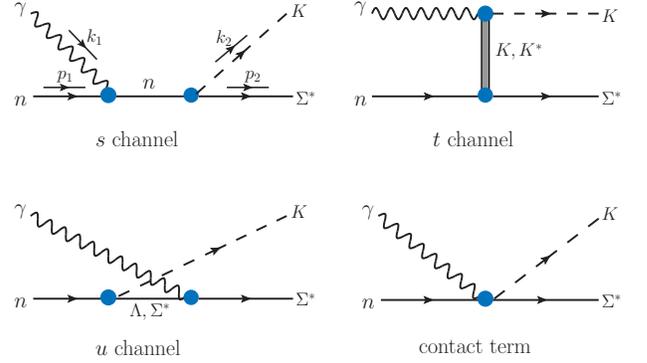}
\end{center}
\caption{(Color online)Feynman diagrams for the $\protect\gamma n\rightarrow
K^{+}\Sigma ^{\ast }(1385)^{-}$ reaction. {(The $K^{\ast }$ $t$-channel
exchange is not included in the present calculation since its contribution
is negligibly small~\protect\cite{yo08,he14}.)}}
\label{Fig:fyd}
\end{figure}

The basic tree-level Feynman diagrams for the $\gamma n\rightarrow
K^{+}\Sigma ^{\ast }(1385)^{-}$ reaction are depicted in Fig.~\ref{Fig:fyd}. 
These include the $s$-channel nucleon pole, the $t$-channel $%
K$ and $K^{\ast }$ exchanges, the $\Lambda $, and $\Sigma ^{\ast }$
intermediate $u$ channel and the contact-type interaction current. In the
present work, the contribution from $t$-channel $K^{\ast }$ exchange is
omitted since it is known to be negligibly small~\cite{yo08,he14}.

\subsection{Lagrangians and amplitudes}

For the $s$ and $t$ channels and the contact term, the relevant effective
Lagrangian densities read as follows~\cite{yo08,gao10,he14},
\begin{align}
\mathcal{L}_{\gamma KK}& =ieA_{\mu }\left[K^{-}(\partial^{\mu}K^{+})-(%
\partial ^{\mu }K^{-})K^{+}\right]~, \\
\mathcal{L}_{KN\Sigma ^{\ast }}& =\frac{f_{KN\Sigma ^{\ast }}}{m_{K}}\bar{N}{%
\Sigma }^{\ast \mu }\cdot \tau (\partial _{\mu }{K})+\text{h.c.}~, \\
\mathcal{L}_{\gamma NN}& =-e\bar{N}\big(Q_{N} A\!\!\!/-\frac{\kappa _{N}}{
4m_{N}}\sigma_{\mu \nu }F^{\mu \nu }\big)N~, \\
\mathcal{L}_{\gamma KN\Sigma ^{\ast }} & = -ie\frac{f_{KN\Sigma ^{\ast }}}{
m_{K}}A^{\mu }\big(\bar{p}\,{\Sigma}_{\mu }^{\ast 0}+\sqrt{2}\bar{n}\,{\Sigma%
}_{\mu }^{\ast -}\big) K^{+}+\text{H.c.}~,
\end{align}%
with the isospin structure of $KN\Sigma ^{\ast }$ coupling given by
\begin{equation}
{\Sigma}\cdot \tau =%
\begin{pmatrix}
{\Sigma}^{0} & \sqrt{2}{\Sigma}^{+} \\
\sqrt{2}{\Sigma}^{-} & -{\Sigma}^{0}%
\end{pmatrix}%
,\quad {K}=\binom{K^{+}}{{K}^{0}}, \quad N=\binom{p}{n},
\end{equation}%
and $F^{\mu \nu }=\partial ^{\mu }A^{\nu }-\partial ^{\nu }A^{\mu }$, where $%
A^{\mu }$, $K$, $\Sigma ^{\ast \mu }$, and $N$ are the photon, kaon, $\Sigma
^{\ast }(1385)$, and nucleon fields. {The kaon and nucleon masses,
respectively, are} $m_{K}$ and $m_{N}${;} $Q_{N}$ is the charge of the
hadron in {units} of $e=\sqrt{4\pi \alpha }$, with $\alpha $ being the
fine-structure constant{, and} $\kappa _{N}=-1.913$ {is the anomalous
magnetic moment} for the neutron \cite{yh13}.

As alluded to in the Introduction, below we will introduce an interpolating
Regge treatment for the $t$ channel. In doing so, the coupling constant $%
f_{KN\Sigma ^{\ast }}$ for this channel will be replaced by a free
parameters $f^{\mathrm{Regge}}_{KN\Sigma ^{\ast }}$ that will be fitted to
experimental data and thus need not be the same as $f_{KN\Sigma ^{\ast }}$
for the $KN\Sigma ^{\ast }$ vertex in the $s$ channel.

For the $u$-channel $\Lambda (1116)$ exchange, the effective Lagrangians for
$\gamma \Lambda \Sigma ^{\ast }$ and $KN\Lambda $ couplings are~\cite{gao10}
\begin{align}
\mathcal{L}_{\gamma \Lambda \Sigma ^{\ast }}& =-\frac{ief_{1}}{2m_{\Lambda }}%
\bar{\Lambda}\gamma _{\nu }\gamma _{5}F^{\mu \nu }\Sigma _{\mu }^{\ast }
\notag \\
&\mbox{}\qquad -\frac{ef_{2}}{(2m_{\Lambda })^{2}}(\partial_{\nu }\bar{%
\Lambda}) \gamma _{5}F^{\mu \nu }\Sigma _{\mu }^{\ast } +\text{H.c.}~, \\
\mathcal{L}_{KN\Lambda }& =-ig_{_{KN\Lambda }}\bar{N}\gamma _{5}\Lambda K+%
\text{H.c.}~,
\end{align}%
where $f_{1}$ and $f_{2}$ are magnetic coupling constants determined from
the partial decay width $\Gamma _{\Sigma ^{\ast }\rightarrow \Lambda \gamma}$%
~\cite{pdg} and model-predicted helicity amplitudes~\cite{mw91}. With the
quark-model result for the helicity amplitudes $A_{1/2}$ and $A_{3/2}$, we
get
\begin{equation}
f_{1}=4.52~,\qquad f_{2}=5.63~.
\end{equation}%
Furthermore, the coupling-constant value $g_{KN\Lambda }=-13.24$ is an
estimate based on flavor SU(3) symmetry relations \cite{gao10,oh06,yo08}.

For the $u$-channel $\Sigma ^{\ast }$ exchange, the effective Lagrangian for
$\gamma \Sigma ^{\ast }\Sigma ^{\ast }$ is~\cite{gao10}
\begin{equation}
\mathcal{L}_{\gamma \Sigma ^{\ast }\Sigma ^{\ast }}=e\bar{\Sigma}_{\mu
}^{\ast }A_{\nu }\Gamma _{\gamma \Sigma ^{\ast }}^{\nu ,\mu \alpha }\Sigma
_{\alpha }^{\ast }~,
\end{equation}%
with%
\begin{align}
A_{\nu }\Gamma _{\gamma \Sigma ^{\ast }}^{\nu ,\mu \alpha }& =Q_{\Sigma
^{\ast }}A_{\nu }\left[ g^{\mu \alpha }\gamma ^{\nu }-\tfrac{1}{2}(\gamma
^{\mu }\gamma ^{\alpha }\gamma ^{\nu }+\gamma ^{\nu }\gamma ^{\mu }\gamma
^{\alpha })\right]  \notag \\
&\mbox{}\qquad -\frac{\kappa _{\Sigma ^{\ast }}}{2m_{N}}\sigma ^{\nu \beta
}\partial _{\beta }A_{\nu }g^{\mu \alpha }~,
\end{align}%
where $Q_{\Sigma ^{\ast }}$ and $\kappa _{\Sigma ^{\ast }}$ denote the
electric charge (in units of $e$) and the anomalous magnetic moment of $%
\Sigma ^{\ast }(1385)$, respectively. Following the quark-model prediction,
we take $\kappa _{\Sigma ^{\ast -}}=-2.43$~\cite{gao10,Lichtenberg:1976fi}.

To account for the internal structure of hadrons, we introduce
phenomenological form factors. For the $s$ and $u$ channels, we adopt the
functional form used in Refs.~\cite{he14,gao10}, i.e.,
\begin{equation}
F_{s/u}(q_{ex}^{2})=\frac{\Lambda _{s/u}^{4}}{\Lambda
_{s/u}^{4}+(q_{ex}^{2}-m_{ex}^{2})^{2}}~,
\end{equation}%
and for the $t$-channel $K$ exchange, we take the monopole form
\begin{equation}
F_{t}(q_{ex}^{2})=\frac{\Lambda _{t}^{2}-m_{ex}^{2}}{\Lambda
_{t}^{2}-q_{ex}^{2}}~,
\end{equation}%
where $q_{ex}$ and $m_{ex}$ are the respective four-momenta and masses of
the exchanged hadrons. The values of the cutoff parameters $\Lambda_{s}$, $%
\Lambda_{u}$, and $\Lambda_{t}$ will be determined here by fits to the data.

With the effective Lagrangian densities as listed above, the invariant
channel scattering amplitudes for the $\gamma n\rightarrow K^{+}\Sigma
^{\ast}(1385)^{-}$ reaction are given as
\begin{equation}
-i\mathcal{M}_{x}=\bar{u}_{\mu }(p_{2},\lambda _{\Sigma ^{\ast }})A_{x}^{\mu
\nu }u(p_{1},\lambda _{n})\epsilon _{\nu }(k_{1},\lambda _{\gamma })~,
\end{equation}%
where the index $x=s,u,t$ corresponds to the appropriate Mandelstam
variable, and $x=c$ denotes the contact-term contribution; the photon
polarization vector is $\epsilon $, and $u_{\mu}$ and $u$ are dimensionless
Rarita-Schwinger and Dirac spinors, respectively; $\lambda _{\Sigma ^{\ast }}
$, $\lambda _{n}$ and $\lambda _{\gamma }$ are the helicities for the $%
\Sigma ^{\ast }(1385)$, the neutron, and the photon, respectively. The
four-momentum dependence here can be read off of Fig.~\ref{Fig:fyd}.

The reduced $A_{x}^{\mu \nu }$ amplitudes for $s$-, $t$-, and $u$-channel
contributions read
\begin{align}
A_{s}^{\mu \nu }& =-\sqrt{2}\frac{ef_{KN\Sigma ^{\ast }}}{2m_{K}m_{N}}\frac{%
\kappa _{N}}{s-m_{N}^{2}} k_{2}^{\mu }(\rlap{$\slash$}k_{1}+\rlap{$\slash$}%
p_{1}+m_{N})\gamma ^{\nu }\rlap{$\slash$}k_{1}F_{s}, \\
A_{t}^{\mu \nu }& =\sqrt{2}\frac{ef_{KN\Sigma ^{\ast }}}{m_{K}}\frac{1}{%
t-m_{K}^{2}}(k_{2}^{\nu }-q_{t}^{\nu })q_{t}^{\mu }F_{t}~,  \label{eq:A_t} \\
A_{u,\Lambda }^{\mu \nu }& =g_{_{KN\Lambda }}\Big\{\frac{ef_{1}}{2m_{\Lambda
}}\gamma _{5}\left( k_{1}^{\mu }\gamma ^{\nu }-g^{\mu \nu }\rlap{$\slash$}%
k_{1}\right)  \notag \\
&\mbox{}\quad +\frac{ef_{2}}{(2m_{\Lambda })^{2}}\gamma _{5}\left(
k_{1}^{\mu }q_{u}^{\nu }-g^{\mu \nu }k_{1}\cdot q_{u}\right) \Big\} \frac{%
\rlap{$\slash$}q_{u}+m_{\Lambda }}{u-m_{\Lambda }^{2}}\gamma _{5}F_{u}, \\
A_{u,\Sigma ^{\ast }}^{\mu \nu }& =\sqrt{2}\frac{Q_{\Sigma ^{\ast
}}ef_{KN\Sigma ^{\ast }}}{m_{K}} \Big\{\big[g^{\mu \alpha }\gamma ^{\nu } -%
\tfrac{1}{2} (\gamma ^{\mu }\gamma ^{\alpha }\gamma ^{\nu }+\gamma ^{\nu
}\gamma ^{\mu }\gamma ^{\alpha })\big]  \notag \\
&\mbox{}\quad \quad -\frac{\kappa _{\Sigma ^{\ast }}}{2m_{N}}\sigma ^{\nu
\rho }k_{1\rho }g^{\mu \alpha }\Big\}\frac{\rlap{$\slash$}q_{u}+m_{\Sigma
^{\ast }}}{u-m_{\Sigma ^{\ast }}^{2}}G_{ \alpha \beta} k_{2}^{\beta} F_{u},
\end{align}%
with%
\begin{align}
G_{\alpha \beta }& =g_{\alpha \beta }-\frac{1}{3}\gamma _{\alpha }\gamma
_{\beta }-\frac{2(q_{u})_{\alpha }(q_{u})_{\beta }}{3m_{\Sigma ^{\ast }}^{2}}
\notag \\
&\mbox{}\qquad\qquad -\frac{\gamma _{\alpha }(q_{u})_{\beta }-\gamma _{\beta
}(q_{u})_{\alpha }}{3m_{\Sigma ^{\ast }}}~,
\end{align}%
where $s=q_{s}^{2}=(k_{1}+p_{1})^{2}$, $t=q_{t}^{2}=(k_{1}-k_{2})^{2}$ and $%
u=q_{u}^{2}=(p_{2}-k_{1})^{2}$ are the Mandelstam variables.

The contact-type interaction current amplitude $A_c^{\mu\nu}$ is given in
Sec.~\ref{sec:GI} below.

\subsection{\boldmath Interpolating Reggeized $t$-channel form factor}

\label{sec:inter}

Standard Regge phenomenology for the $t$-channel meson exchange consists of
replacing the product of the form factor and meson propagator in Eq.~(\ref%
{eq:A_t}) according to~\cite{GLV97,Haberzettl:2015exa}
\begin{equation}
\frac{F_t(t)}{t-m_K^2} \to \frac{\mF_t(t)}{t-m_K^2}~,  \label{eq:tRegge}
\end{equation}
where the residual Regge function $\mF_t$ contains all higher-mass poles
along the Regge trajectory above the base state at $t=m_K^2$. The
Reggeization of the $t$ channel thus effectively corresponds to a
prescription of how to choose the corresponding form factor.

Using the notation of Ref.~\cite{Haberzettl:2015exa}, the residual function
for the present application is written as
\begin{equation}
\mF_t(t) =\left(\frac{s}{s_\scale}\right)^{\alpha_K(t)} \frac{N_K\big(%
\alpha_K(t);\eta\big) }{\Gamma\big(1+\alpha _{K}(t)\big)} \frac{\pi \alpha
_{K}(t)}{\sin \big(\pi \alpha _{K}(t)\big)}~,
\end{equation}
where
\begin{equation}
\alpha_K(t) = \alpha^{\prime }_K (t- M_K^2)
\end{equation}
is the kaon trajectory with the usual slope parameter $\alpha^{\prime
}_K=0.7\,\text{GeV}^{-2}$~\cite{he14,GLV97,tc07}. The scale parameter
of the exponential factor is taken as $s_\scale= 1\,\text{GeV}^{2}$. The
signature function is given as~\cite{Haberzettl:2015exa}
\begin{equation}
N[\alpha_K(t);\eta] = \eta + (1-\eta)e^{-i\pi\alpha_K(t)}~,
\label{eq:fitphase}
\end{equation}
where $\eta$ is a (phenomenological) real parameter whose three standard
values are
\begin{equation}
\eta =%
\begin{cases}
\frac{1}{2}~, & \text{pure-signature trajectories}~, \\
0~, & \text{add trajectories: rotating phase}~, \\
1~, & \text{subtract trajectories: constant phase}~.%
\end{cases}%
\end{equation}
Without going into details here (for a discussion of this parametrization,
see Ref.~\cite{Haberzettl:2015exa}), only the latter two choices ($\eta=0,1$%
) apply here in view of the degeneracy of the kaon trajectory starting at $%
m_K=495$\,MeV~\cite{GLV97,tc07}. Numerical tests show that for the present
case the best results are produced by the choice
\begin{equation}
\eta=1 \quad\Rightarrow\quad N_K\big(\alpha_K(t);1\big) =1~.
\end{equation}
This corresponds to subtraction of the degenerate secondary trajectory
[starting at $K_1(1270)$] from the primary one. Note that this subtraction
is consistent with our choice of monopole form factor $F_t$ for the standard
Feynman-type single-meson exchange for the $t$-channel since
\begin{equation}
\frac{1}{t-m_K^2}\frac{\Lambda_t^2-m_K^2}{\Lambda_t^2-t} =\frac{1}{t-m_K^2}-%
\frac{1}{t-\Lambda^2_t}~,
\end{equation}
where the secondary pole contribution with `mass' $\Lambda_t$ is also
subtracted. (If the cutoff-mass in this Pauli-Villars-type regularization~%
\cite{PV49} were taken as $\Lambda_t=1.29$\,GeV, this would correspond
\textit{exactly} to the second pole along the degenerate Regge trajectory.
In the present application, however, $\Lambda_t$ is fitted to the data.)

The onset of the `Regge regime' is oftentimes very much under debate in
practical applications, in particular, if Regge exchanges are employed in
medium-energy ranges relevant for baryon-resonance physics. It seems
reasonable, therefore, to consider mechanisms for smooth transitions into
that regime that can be fine-tuned to the requirements of particular
applications~\cite%
{Toki,NK2010,NY11,He:2012ud,NG13,he14,He:2014gga,Wang:2014jxb}. Fitting the
parameters of such an interpolation scheme to experimental data lets the
data `decide' whether Regge exchanges should be necessary or not for a
particular process at a particular photon energy.

Since Regge phenomenology applies to high $s$ and low $|t|$, we adopt here
the interpolating mechanism proposed in Ref.~\cite{NK2010} that provides
separate switching functions for $t$ and $s$ which we write as
\begin{align}
R_{s}(s)=\frac{1}{1+e^{-(s-s_{R})/s_{0}}}, \quad R_{t}(t) =\frac{1}{%
1+e^{-(t+t_{R})/s_{0}}}~,  \label{eq:RsRt}
\end{align}
where $s_R$ and $t_R$ describe the centroid values for the transition from
non-Regge to Regge regimes, with $s_0$ and $t_0$ providing the respective
widths of the transition regions. (The sign change for $t_R$ is chosen
merely for convenience to have positive values for both $s_R$ and $t_R$ in
the physical region.) Combined as
\begin{equation}
R(t) = R_s(s)\,R_t(t)~,
\end{equation}
this product provides an interpolating function in $t$ for $s$ fixed by
experiment. The four parameters of this function will be fitted to the
experimental data.

The interpolated Reggeized form factor can then be written as
\begin{equation}
F_t \to \mF_{R,t}(t) = \mF_t(t)\, R(t) + F_t(t)\,\left[1-R(t)\right]~,
\label{eq:Reggeizeft}
\end{equation}
which replaces $F_t$ on the left-hand side of Eq. (\ref{eq:tRegge}), with $\mF%
_{R,t}$, thus providing a smooth interpolation between the usual Feynman case ($R=0
$) and the full Regge case given by the right-hand side (for $R=1$).

\begin{figure}[t!]
\centering
\includegraphics[width=.95\columnwidth,clip=]{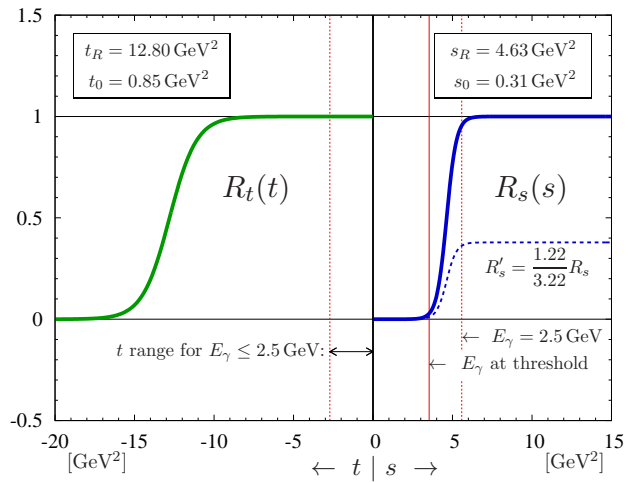}
\caption{(Color online) Interpolating switching functions of Eq.~(\protect
\ref{eq:RsRt}), with parameter values resulting from the present fits (see
Sec.~\protect\ref{sec:results}). The function $R_t(t)$ on the left is
effectively unity for photon laboratory energies $E_\protect\gamma\le 2.5$\,GeV.
The function $R_s(s)$ on the right changes from essentially zero at
threshold to almost unity across the energy range of the present data. For
the correct interpretation of this finding and, in particular, the meaning
of the dotted curve labeled $R^{\prime }_s$ on the right, see text.}
\label{fig:Rswitch}
\end{figure}

Since $s$ is fixed, the switch $R_s$ will only contribute a constant factor.
Hence, the more relevant switch for reproducing detailed features of an
experiment is $R_t$ since it directly affects the description of angular
behavior. While this may offer valuable flexibility for data that show rapid
dependence on the scattering angle, this turns out to be not necessary for
the present application. In fact we will find below that effectively the
fitted values of $t_R$ and $t_0$ correspond to $R_t=1$ across the range of
data considered here (see Fig.~\ref{fig:Rswitch}). For the present
application, therefore, only the switch $R_s$ will matter.

Finally, it is obvious that the interpolation (\ref{eq:Reggeizeft}) does not
change the normalization of the form factor, i.e.,
\begin{equation}
\mF_{R,t}(m_K^2) =1~,  \label{eq:FRnorm}
\end{equation}
which is a necessary condition for the gauge-invariance-preserving procedure
explained subsequently to work.

\subsection{Preserving local gauge invariance}

\label{sec:GI}

As is well known~\cite{Haberzettl:2015exa,GLV97}, Reggeization of the $t$%
-channel exchange destroys gauge-invariance of the production current.
However, following Ref.~\cite{Haberzettl:2015exa}, this can easily be
restored by generalizing the gauge-invariance preserving procedure of Ref.~%
\cite{hh06} to the Regge case. Imposing \textit{local} gauge invariance in
the form of generalized Ward-Takahashi identities, this results in the
contact-type interaction current~\cite{Haberzettl:2015exa}
\begin{equation}
A^{\mu\nu}_c=e \sqrt{2}\frac{f_{KN\Sigma ^{\ast }}}{m_{K}}\big[g^{\mu \nu }%
\mF_{R,t}(t)-k_{2}^{\mu }C^{\nu }\big]~,  \label{eq:A_c}
\end{equation}%
with
\begin{align}
C^{\nu }&=-(2k_{2}-k_{1})^{\nu }\frac{\mF_{R,t}-1}{t-m_{K}^{2}}F_{u}  \notag
\\
&\qquad\mbox{} +(2p_{2}-k_{1})^{\nu }\frac{F_{u}-1}{u-m_{\Sigma ^{\ast }}^{2}%
}\mF_{R,t}  \notag \\
& \mbox{}\qquad +\hat{A}(1-F_{t})(1-F_{u})  \notag \\
&\qquad\qquad\mbox{}\times \left[ \frac{(2k_{2}-k_{1})^{\nu }}{t-m_{K}^{2}}-%
\frac{(2p_{2}-k_{1})^{\nu }}{u-m_{\Sigma ^{\ast }}^{2}}\right] .
\label{eq:Caux}
\end{align}
In view of the normalization (\ref{eq:FRnorm}), the auxiliary current $C^\nu$
is manifestly nonsingular at the primary $t$-channel pole for $t=m_K^2$,
however, it still retains the high-lying poles along the Regge trajectory
via $\mF_{R,t}$. The latter singularities are necessary to cancel the
corresponding gauge-invariance-violating contributions of the
production-current four-divergence resulting from Reggeization~\cite%
{Haberzettl:2015exa}.

The last $\hat{A}$-dependent term in (\ref{eq:Caux}) is manifestly
transverse and nonsingular. The function $\hat{A}=\hat{A}(t,u)$ here is a
Lorentz-covariant, crossing-symmetric phenomenological function that must
vanish at high energies, but otherwise can be freely chosen to improve fits
to the data. Note here that the preceding Eq.~(\ref{eq:Caux}) follows from
Eq. (31) of Ref.~\cite{hh06} by choosing the function $\hat{h}$ appearing
there as $\hat{h}=1-\hat{A}$. The vanishing high-energy limit of $\hat{A}$
is necessary to prevent the \textquotedblleft violation of scaling
behavior\textquotedblright\ noted in Ref.~\cite{DrellLee1972} if $\hat{h}$
is different from unity at high energies. We simply choose here
\begin{equation}
\hat{A}(t,u)=A_{0}\frac{\Lambda _{c}^{4}}{\Lambda _{c}^{4}+(s-s_\threshold%
)^{2}}~,
\end{equation}%
with%
\begin{equation}
s_\threshold=(m_{\Sigma }+m_{K})^{2}~,
\end{equation}%
which has the (dimensionless) value $A_{0}$ at the reaction threshold $s=s_%
\threshold$. This choice has two parameters, the strength $A_{0}$ and cutoff
$\Lambda _{c}$. For simplicity, we take $\Lambda _{c}=3$ GeV and use only $%
A_{0}$ as a fit parameter. (There is no particular reason for choosing this
cutoff value, other than not having {$\hat{A}$} fall off too rapidly for the
present energy range.)

\section{Results and discussion}

\label{sec:results}

The unpolarized differential cross section for the $\gamma n\rightarrow
K^{+}\Sigma ^{\ast}(1385)^{-}$ reaction at the center of mass (c.m.) frame
is given by
\begin{equation}
\frac{d\sigma }{d\cos \theta }=\frac{1}{32\pi s}\frac{\left\vert \vec{k}%
_{2}^{{~\mathrm{c.m.}}}\right\vert }{\left\vert \vec{k}_{1}^{{~\mathrm{c.m.}}%
}\right\vert }\left( \frac{1}{4}\sum\limits_{\lambda }\left\vert \mathcal{M}%
\right\vert ^{2}\right)
\end{equation}%
where $s=(k_{1}+p_{1})^{2}\equiv W^{2}$ with $W$ being the total energy and $%
\theta $ denotes the angle of the outgoing $K^{+}$ meson relative to beam
direction in the c.m. frame, while $\vec{k}_{1}^{{~\mathrm{c.m.}}}$ and $%
\vec{k}_{2}^{{~\mathrm{c.m.}}}$ are the three-momenta of initial photon beam
and final kaon meson, respectively.

\subsection{Fitting procedure}

As discussed in the Introduction and Sec.~\ref{sec:formalism}, we consider
here only the ``background" contributions, namely, the $s$ channel with
nucleon-pole exchange, the Reggeized $t$ channel with $K$ exchange, the $u$
channel with $\Lambda (1116) $ and $\Sigma ^{\ast }(1385)$ exchanges and the
contact term for the $\gamma n\rightarrow K^{+}\Sigma ^{\ast }(1385)^{-}$
process. The formalism was presented in the previous section, and the relevant input parameters  are collected in Table \ref{table:Lparameters}.
\renewcommand\tabcolsep{0.22cm}
\renewcommand{\arraystretch}{1.8}
\begin{table}[t!]  
\caption{Input parameters for the formalism used in this work.}
\begin{tabular}{|c|c|c|c|c|}
\hline
$f_1$ & $f_2$ & $g_{KN\Lambda}$ &$\kappa_n$& $\kappa_{\Sigma^{*-}}$ \\ \hline
$4.52$ & $5.63$ & $-13.24$ &-1.91& $-2.43$\\ \hline\hline
 $\alpha'_K$~[GeV$^{-2}$]&$s_{sc}$~[GeV$^2$] & $\eta$ & $\Lambda _{c}$~[GeV]&\\ \hline
 0.7&1.0& 1.0 & 3.0 &\\
\hline
\end{tabular}%
\label{table:Lparameters}
\end{table}

\begin{figure}[t!]
\centering
\includegraphics[width=.95\columnwidth]{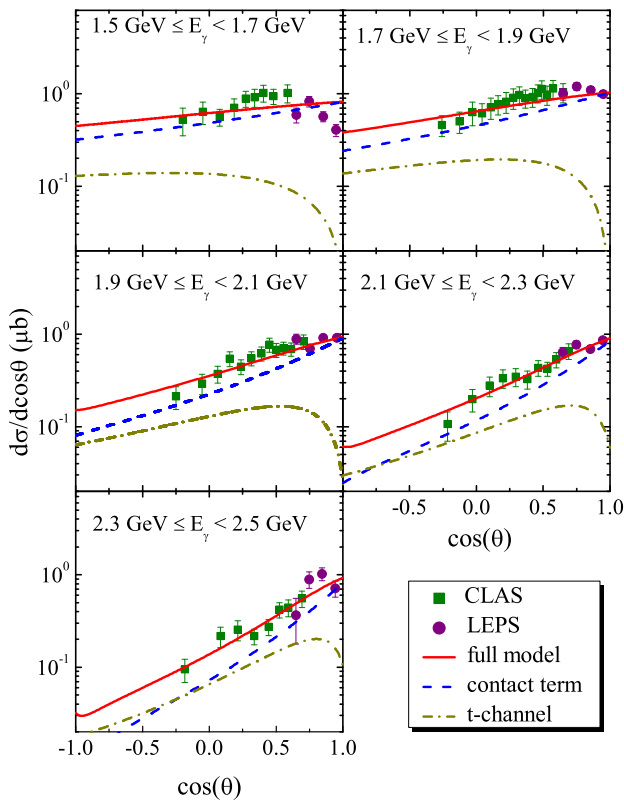}
\caption{(Color online) Differential cross section $d\protect\sigma /d\cos
\protect\theta $ for $\Sigma ^{\ast }(1385)$ photoproduction off a neutron
as function of $\cos\protect\theta$. Data are from \protect\cite%
{clas14,leps09}.}
\label{Fig: dcs}
\end{figure}

The preliminary CLAS data \cite{clas14} and LEPS data \cite{leps09}
will be fitted with the help of the \textsc{minuit} code in the \textsc{%
cernlib}. In this work, we minimize $\chi^2$ per degree of freedom (dof) for
the differential cross sections ${d\sigma }/{d\cos \theta }$ for the CLAS
and LEPS data by fitting the nine parameters $f_{KN\Sigma ^{\ast }}^{\text{%
Regge}}$, $s_{R}$, $s_{0}$, $t_{R}$, $t_{0}$, $\Lambda_{s}$, $\Lambda_{u}$, $%
\Lambda_{t}$, and $A_{0}$ using a total of 75 data points as displayed in
Fig.~\ref{Fig: dcs}. The differential cross-section data are given for five
intervals of the beam energy $E_{\gamma }$ from 1.5 GeV up to 2.5 GeV.

The parameter values determined in this manner are given in Table~\ref%
{table:parameters}, with an reduced value $\chi ^{2}/\text{dof}=1.75$, which
suggests the CLAS~\cite{clas14} and LEPS~\cite{leps09} data sets can indeed
can be reproduced quite well by the presently considered mechanisms, without
any need for explicit intermediate $s$-channel resonances.
\renewcommand\tabcolsep{0.03cm}
\renewcommand{\arraystretch}{2}
\begin{table}[b!]
\caption{Fitted values of free parameters and corresponding reduced $\protect\chi %
^{2}/\text{dof}$ value.}
\begin{tabular}{|c|c|c|c|c|}
\hline
~$f_{KN\Sigma ^{\ast }}^{\text{Regge}}$~\ystrut{-1.5ex\ystrut{3ex}} & ~$s_{R}
$~[GeV$^{2}$]~ & ~$s_{0}$~[GeV$^{2}$]~ & ~$t_{R}$~[GeV$^{2}$]~ & ~$t_{0}$%
~[GeV$^{2}$]~ \\ \hline
$-1.22\pm 0.02$ & 4.63$\pm 0.06$ & 0.31$\pm 0.01$ & 12.80$\pm 0.35$ & 0.85$%
\pm 0.21$ \\ \hline\hline
$A_{0}$\ystrut{-1ex}\ystrut{2.5ex} & $\Lambda _{s}$~[GeV] & $\Lambda _{u}$%
~[GeV] & $\Lambda _{t}$~[GeV] & $\chi ^{2}/{\text{dof}}$ \\ \hline
0.03$\pm 0.01$ & 1.03$\pm 0.12$ & 0.81$\pm 0.03$ & 1.45$\pm 0.05$ & 1.75 \\
\hline
\end{tabular}%
\label{table:parameters}
\end{table}

This fit quality is achieved with reasonable cutoff values $\Lambda_x$ ($%
x=s,u,t$) around the usual empirical 1-GeV value. The Regge interpolation
parameters $t_R$ and $t_0$ are largely irrelevant since, as shown in Fig.~%
\ref{fig:Rswitch}, the function $R_t(t)=1$ across the range of data
considered here.

The functional behavior of $R_s(s)$ for the parameters $s_R$ and $s_0$, on
the other hand, exhibits a rapid variation from close to zero at threshold
to almost unity at the upper energy end, $E_\gamma=2.5$\,GeV, of data
employed here. While this seems to suggest that Regge behavior is fully
switched on at $E_\gamma=2.5$\,GeV, this has to be taken with some caution
because across the same energy range, the fitted value $f_{KN\Sigma ^{\ast
}}^{\text{Regge}}$ of the Reggeized $t$-channel $KN\Sigma ^{\ast}$ coupling
strength drops in magnitude by almost one third, from 3.22 to 1.22. Hence,
since the strength of the Regge contribution is determined only by the
product $f_{KN\Sigma ^{\ast }}^{\text{Regge}}\,R_s(s)$, one cannot really
say at what energy Regge behavior will be switched on fully. Rescaling $%
R_s(s)$ with the ratio of the fitted coupling strength and its orginal SU(3)
value, i.e., $R^{\prime }_s(s)=(1.22/3.22)R_s(s)$ depicted as the dotted
curve in Fig.~\ref{fig:Rswitch}, one can say, however, that the Regge
\textit{vs.} non-Regge contribution must lie somewhere in the region between
the solid $R_s$ and the dotted $R^{\prime }_s$ curves in Fig.~\ref%
{fig:Rswitch}. This means, in particular, that Regge behavior already plays
a significant role in this energy range, even if the detailed changeover
behavior cannot be pinned down precisely by the present approach.

\subsection{\boldmath Cross section for $\protect\gamma n\rightarrow
K^{+}\Sigma ^{\ast }(1385)^{-}$}

As can be seen from Fig.~\ref{Fig: dcs}, the differential cross section $%
d\sigma/d\Omega$ for both CLAS and LEPS data sets~\cite{clas14,leps09} are
well reproduced in our model, in particular, at the high-energy end of the
data range. At lower energies, some structure at forward angles shown by the
LEPS data is not so well described and, if it could be corroborated by other
independent experiments, may require a more sophisticated approach,
including perhaps some $s$-channel resonances. With the present simple
``background" model, however, the $s$ and $u$ channels contribute so little
to the cross section that we have omitted showing these very small
contributions in the figure.

The bulk of the contributions are seen to come from the Reggeized $t$
channel and, in particular, from the contact term of Eq.~(\ref{eq:A_c}).
Since the contact current can be understood as the minimal contribution from
the hadronic final-state interaction (FSI) necessary to preserve gauge
invariance~\cite{hh06,hh97}, this may indicate that $K\Sigma$ FSI may play a
role if one cannot resolve the discrepancies here by other means. However,
since there are no data to constrain the parameters of such an FSI, an
actual reliable calculation of such FSI processes would be impossible at
present.

\begin{figure}[b!]
\centering
\includegraphics[width=1.1\columnwidth]{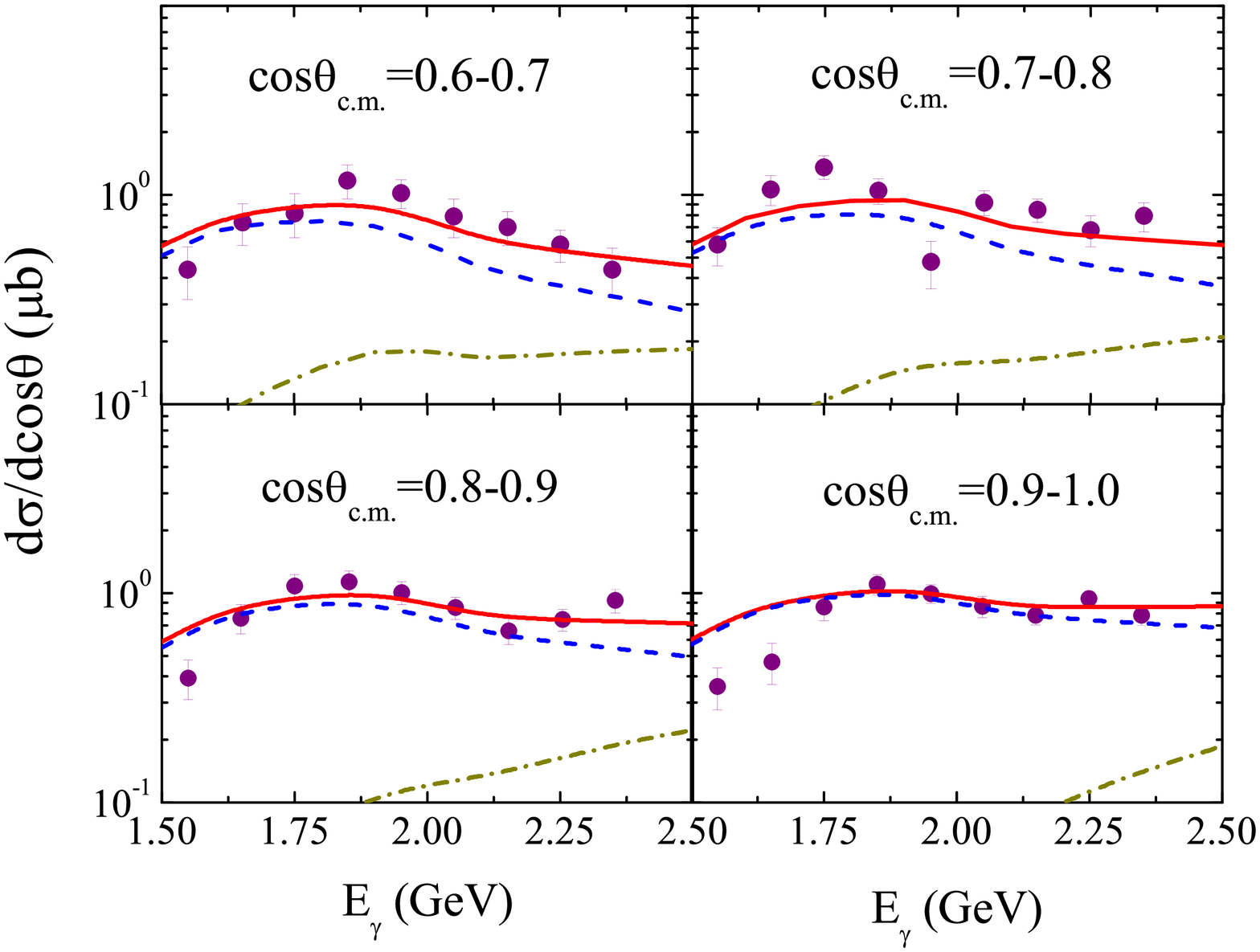}
\caption{(Color online) Differential cross section $d\protect\sigma /d\cos
\protect\theta $ for $\Sigma ^{\ast }(1385)$ photoproduction off a neutron
plotted against photon energy $E_{\protect\gamma }$ for LEPS data
\protect\cite{leps09}. The line styles here are the same as in Fig. 3}
\label{Fig: LEPS}
\end{figure}

\begin{figure}[t!]
\begin{center}
\includegraphics[width=.95\columnwidth]{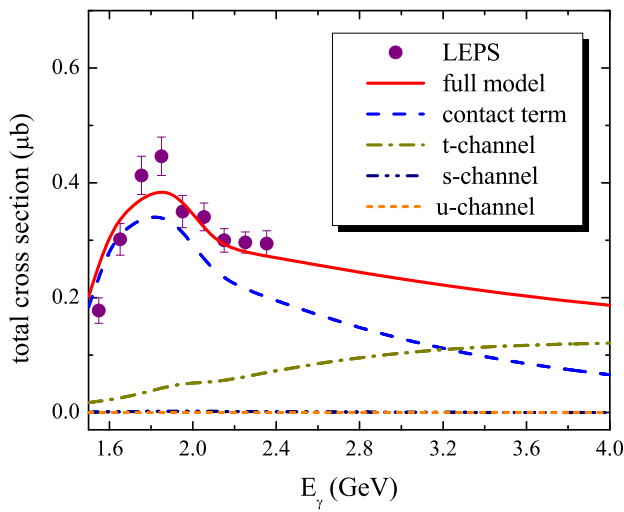}
\end{center}
\caption{(Color online) Total cross section for $\Sigma ^{\ast }(1385)$
photoproduction off a neutron, compared with LEPS data \protect\cite{leps09}.
}
\label{Fig:total}
\end{figure}

The primary objective of the present investigation was the description of
the preliminary CLAS data~\cite{clas14}. However, since they still have
large uncertainties, we also wanted to test our model for LEPS data~\cite%
{leps09}. Using the fit parameters of Table~\ref{table:parameters}, we see
that we can reproduce reasonably well the differential cross sections, Fig,~%
\ref{Fig: LEPS}, and the total cross sections, Fig.~\ref{Fig:total}, of that
experiment. The LEPS data can be reproduced in our model except for some discrepancies at low energies, which is similar to the results in Ref.~\cite{Oh:2010ur} obtained with the formalism in Ref.~\cite{yo08}. From Fig.~\ref{Fig:total}, we again find a very large part of the final
result is determined by the $t$ channel and, again, by the
gauge-invariance-preserving contact term. The $s$- and $u$-channel
contributions are negligibly small.

\section{Summary}

\label{sec:summary}

We have presented here an effective Lagrangian approach to the
photoproduction reaction $\gamma n\to K^+\Sigma^*(1385)^-$ with a Reggeized $%
t$-channel exchange that permits smooth interpolation between standard
Feynman-type single-meson exchange and a full-fledged Regge trajectory
exchange. The present work is the first application of the method put
forward recently by the present authors~\cite{Haberzettl:2015exa} that
preserves full \textit{local} gauge invariance in terms of generalized
Ward-Takahashi identities~\cite{Kazes1959,hh97}.

Applying the model to recent CLAS data~\cite{clas14} and somewhat older LEPS
data~\cite{leps09}, we find good agreement with differential and total cross
sections, with $\chi^2/\text{dof}=1.75$. We cannot fit the data very well
using the usual Feynman-type $t$-channel exchange alone. Inclusion of Regge
trajectories is essential to achieve the fit quality exhibited in Figs.~\ref%
{Fig: dcs}, \ref{Fig: LEPS}, and \ref{Fig:total}, in particular, with the
added flexibility of the Reggeized interpolating $t$-channel form factor.
Equally important are the contributions from the contact-type interaction
current term that results from the gauge-invariance-preserving procedure of
Ref.~\cite{Haberzettl:2015exa} since they account for a large part of the cross
sections. The \textit{microscopically} correct treatment of \textit{local}
gauge invariance thus turns out to be an essential ingredient of the present
model.

As argued, the dominance of the contact term may point to $K\Sigma$
final-state contributions being important for this process; however, there
are no data that would constrain any calculation along those lines, thus
unfortunately making this an untestable proposition (at least, at present)

In summary, since the present model does indeed reproduce quite well the
main features of the process $\gamma n\to K^+\Sigma^*(1385)^-$, we are
confident that the mechanisms incorporated in the model provide the dominant
physics of the reaction, in particular, that the simplified duality
treatment where $s$-channel baryon resonances are traded for $t$-channel
meson Regge trajectories is indeed capable of describing magnitudes and
average features of the observables.

To describe more detailed structures of the cross sections, inclusion of $s$%
-channel resonances may very well be necessary. However, to warrant
expanding efforts in this direction, more precise data for $\Sigma ^{\ast
}(1385)$ production are necessary, covering wider energy and angle ranges.
Such experiments could be carried out at JLab (CLAS) or CERN (COMPASS).

\acknowledgements

One of the authors (X.Y.W.) is grateful to Dr. Hao Xu for communication of
the \textsc{minuit} code. This project is partially supported by the Major State
Basic Research Development Program in China under grant 2014CB845405 and the
National Natural Science Foundation of China under grant 11275235.

\end{document}